\documentstyle[12pt]{article}

\author{Alexander Vilenkin and Serge Winitzki\\
\\
Institute of Cosmology\\
Dept. of Physics and Astronomy\\
Tufts University\\
Medford, MA 02155}
\title{Detection of Particles Under Potential Barrier
}
\date{\relax
}

\def\NEG#1{{\rlap/#1}}

\catcode`\@=11

\@addtoreset{equation}{section}

\@addtoreset{figure}{section}
\def\thefigure{\thesection.\@arabic\c@figure}

\@addtoreset{table}{section}
\def\thetable{\thesection.\@arabic\c@table}

\let\DOTSI\relax
\def\RIfM@{\relax\ifmmode}
\def\FN@{\futurelet\next}
\newcount\intno@
\def\iint{\DOTSI\intno@\tw@\FN@\ints@}
\def\ints@{\findlimits@\ints@@}
\newif\iflimtoken@
\newif\iflimits@
\def\findlimits@{\limtoken@true\ifx\next\limits\limits@true
 \else\ifx\next\nolimits\limits@false\else
 \limtoken@false\ifx\ilimits@\nolimits\limits@false\else
 \ifinner\limits@false\else\limits@true\fi\fi\fi\fi}
\def\multint@{\int\ifnum\intno@=\z@\intdots@                                
 \else\intkern@\fi                                                          
 \ifnum\intno@>\tw@\int\intkern@\fi                                         
 \ifnum\intno@>\thr@@\int\intkern@\fi                                       
 \int}                                                                      
\def\multintlimits@{\intop\ifnum\intno@=\z@\intdots@\else\intkern@\fi
 \ifnum\intno@>\tw@\intop\intkern@\fi
 \ifnum\intno@>\thr@@\intop\intkern@\fi\intop}
\def\intic@{\mathchoice{\hskip.5em}{\hskip.4em}{\hskip.4em}{\hskip.4em}}
\def\negintic@{\mathchoice
 {\hskip-.5em}{\hskip-.4em}{\hskip-.4em}{\hskip-.4em}}
\def\ints@@{\iflimtoken@                                                    
 \def\ints@@@{\iflimits@\negintic@\mathop{\intic@\multintlimits@}\limits    
  \else\multint@\nolimits\fi                                                
  \eat@}                                                                    
 \else                                                                      
 \def\ints@@@{\iflimits@\negintic@
  \mathop{\intic@\multintlimits@}\limits\else
  \multint@\nolimits\fi}\fi\ints@@@}
\def\intkern@{\mathchoice{\!\!\!}{\!\!}{\!\!}{\!\!}}
\def\plaincdots@{\mathinner{\cdotp\cdotp\cdotp}}
\def\intdots@{\mathchoice{\plaincdots@}
 {{\cdotp}\mkern1.5mu{\cdotp}\mkern1.5mu{\cdotp}}
 {{\cdotp}\mkern1mu{\cdotp}\mkern1mu{\cdotp}}
 {{\cdotp}\mkern1mu{\cdotp}\mkern1mu{\cdotp}}}

\newif\iffirstchoice@
\firstchoice@true
\def\textfonti{\the\textfont\@ne}
\def\textfontii{\the\textfont\tw@}
\def\text{\RIfM@\expandafter\text@\else\expandafter\text@@\fi}
\def\text@@#1{\leavevmode\hbox{#1}}
\def\text@#1{\mathchoice
 {\hbox{\everymath{\displaystyle}\def\textfonti{\the\textfont\@ne}%
  \def\textfontii{\the\textfont\tw@}\textdef@@ T#1}}
 {\hbox{\firstchoice@false
  \everymath{\textstyle}\def\textfonti{\the\textfont\@ne}%
  \def\textfontii{\the\textfont\tw@}\textdef@@ T#1}}
 {\hbox{\firstchoice@false
  \everymath{\scriptstyle}\def\textfonti{\the\scriptfont\@ne}%
  \def\textfontii{\the\scriptfont\tw@}\textdef@@ S\rm#1}}
 {\hbox{\firstchoice@false
  \everymath{\scriptscriptstyle}\def\textfonti
  {\the\scriptscriptfont\@ne}%
  \def\textfontii{\the\scriptscriptfont\tw@}\textdef@@ s\rm#1}}}
\def\textdef@@#1{\textdef@#1\rm\textdef@#1\bf\textdef@#1\sl\textdef@#1\it}
\def\DN@{\def\next@}
\def\eat@#1{}
\def\textdef@#1#2{%
 \DN@{\csname\expandafter\eat@\string#2fam\endcsname}%
 \if S#1\edef#2{\the\scriptfont\next@\relax}%
 \else\if s#1\edef#2{\the\scriptscriptfont\next@\relax}%
 \else\edef#2{\the\textfont\next@\relax}\fi\fi}

\def\Let@{\relax\iffalse{\fi\let\\=\cr\iffalse}\fi}
\def\vspace@{\def\vspace##1{\crcr\noalign{\vskip##1\relax}}}
\def\multilimits@{\bgroup\vspace@\Let@
 \baselineskip\fontdimen10 \scriptfont\tw@
 \advance\baselineskip\fontdimen12 \scriptfont\tw@
 \lineskip\thr@@\fontdimen8 \scriptfont\thr@@
 \lineskiplimit\lineskip
 \vbox\bgroup\ialign\bgroup\hfil$\m@th\scriptstyle{##}$\hfil\crcr}
\def\Sb{_\multilimits@}
\def\endSb{\crcr\egroup\egroup\egroup}
\def\Sp{^\multilimits@}

\newdimen\ex@
\def\rightarrowfill@#1{$#1\m@th\mathord-\mkern-6mu\cleaders
 \hbox{$#1\mkern-2mu\mathord-\mkern-2mu$}\hfill
 \mkern-6mu\mathord\rightarrow$}
\def\leftarrowfill@#1{$#1\m@th\mathord\leftarrow\mkern-6mu\cleaders
 \hbox{$#1\mkern-2mu\mathord-\mkern-2mu$}\hfill\mkern-6mu\mathord-$}
\def\leftrightarrowfill@#1{$#1\m@th\mathord\leftarrow\mkern-6mu\cleaders
 \hbox{$#1\mkern-2mu\mathord-\mkern-2mu$}\hfill
 \mkern-6mu\mathord\rightarrow$}
\def\overrightarrow{\mathpalette\overrightarrow@}
\def\overrightarrow@#1#2{\vbox{\ialign{##\crcr\rightarrowfill@#1\crcr
 \noalign{\kern-\ex@\nointerlineskip}$\m@th\hfil#1#2\hfil$\crcr}}}

\def\overleftarrow{\mathpalette\overleftarrow@}
\def\overleftarrow@#1#2{\vbox{\ialign{##\crcr\leftarrowfill@#1\crcr
 \noalign{\kern-\ex@\nointerlineskip}$\m@th\hfil#1#2\hfil$\crcr}}}
\def\overleftrightarrow{\mathpalette\overleftrightarrow@}
\def\overleftrightarrow@#1#2{\vbox{\ialign{##\crcr\leftrightarrowfill@#1\crcr
 \noalign{\kern-\ex@\nointerlineskip}$\m@th\hfil#1#2\hfil$\crcr}}}
\def\underrightarrow{\mathpalette\underrightarrow@}
\def\underrightarrow@#1#2{\vtop{\ialign{##\crcr$\m@th\hfil#1#2\hfil$\crcr
 \noalign{\nointerlineskip}\rightarrowfill@#1\crcr}}}

\def\underleftarrow{\mathpalette\underleftarrow@}
\def\underleftarrow@#1#2{\vtop{\ialign{##\crcr$\m@th\hfil#1#2\hfil$\crcr
 \noalign{\nointerlineskip}\leftarrowfill@#1\crcr}}}
\def\underleftrightarrow{\mathpalette\underleftrightarrow@}
\def\underleftrightarrow@#1#2{\vtop{\ialign{##\crcr$\m@th\hfil#1#2\hfil$\crcr
 \noalign{\nointerlineskip}\leftrightarrowfill@#1\crcr}}}

\catcode`\@=\active

\def\frac#1#2{{#1 \over #2}}

\catcode`\@=11

\long\def\QQQ#1#2{}
\def\QTP#1{}
\long\def\QQA#1#2{}

\long\def\TeXButton#1#2{#2}
\def\EXPAND#1[#2]#3{}
\def\NOEXPAND#1[#2]#3{}

\def\LaTeXparent#1{}

\@ifundefined{abstract}{%
\def\abstract{\if@twocolumn
\section*{Abstract (Not appropriate in this style!)}
\else \small
\begin{center}
{\bf Abstract\vspace{-.5em}\vspace{0pt}}
\end{center}
\quotation
\fi}}{}
\@ifundefined{endabstract}{%
\def\endabstract{\if@twocolumn\else\endquotation\fi}}{}
\@ifundefined{maketitle}{\def\maketitle#1{}}{}
\@ifundefined{affiliation}{\def\affiliation#1{}}{}
\@ifundefined{proof}{}{}
\@ifundefined{newfield}{\def\newfield#1#2{}}{}
\@ifundefined{chapter}{\def\chapter#1{\par(Chapter head:)#1\par }}{}
\@ifundefined{part}{\def\part#1{\par(Part head:)#1\par }}{}
\@ifundefined{section}{\def\part#1{\par(Section head:)#1\par }}{}
\@ifundefined{subsection}{\def\part#1{\par(Subsection head:)#1\par }}{}
\@ifundefined{subsubsection}{\def\part#1{\par(Subsubsection head:)#1\par }}{}
\@ifundefined{paragraph}{\def\part#1{\par(Subsubsubsection head:)#1\par }}{}
\@ifundefined{subparagraph}{\def\part#1{\par(Subsubsubsubsection head:)#1\par
}}{}

\def\lesim{\stackrel{<}{\scriptstyle\sim}}

\begin{document}

\maketitle
\begin{abstract}
We introduce a model detector which registers the passage of a particle
through the detector location, without substantially perturbing the particle
wave function. (The exact time of passage is not determined in such
measurements.) We then show that our detector can operate in a classically
forbidden region and register particles passing through a certain point
under a potential barrier. We show that it should be possible to observe the
particle's track under the barrier.
\end{abstract}

\section{Introduction}

The question we would like to address in this paper is:\ does a particle
leave a track while tunneling under a potential barrier? Semiclassical
calculations of the tunneling rate are often performed using the Euclidean
under-barrier trajectory of the particle, ${\bf x}\left( \tau \right) $,
where $\tau $ is the Euclidean (imaginary) time \cite{euclideantime}. This
trajectory does not, of course, correspond to any classical motion of the
particle in the real time, but if $\tau $ is treated as a parameter, then $%
{\bf x}\left( \tau \right) $ gives a well-defined spatial path (called the
most probable escape path \cite{escapepath}). The question is whether this
path is just a mathematical artifact, or, in some sense, the particle does
follow this trajectory in the process of tunneling. For instance, if a
potential barrier is set up in a cloud chamber, will the track disappear
when the particle hits the barrier and reappear on the other side, or will
it be continuous?

At first sight, one could think that under-barrier detection is impossible
for energetic reasons. If the particle's wave function is localized in a
small region where the potential energy is $U\left( {\bf x}\right) $, then,
since the kinetic energy is positive, the average energy of the particle
will be greater than $U\left( {\bf x}\right) $. This seems to imply that any
under-barrier position measurement would throw the particle above the
barrier. This conclusion is indeed correct if the position is measured at a
specific moment of time, which leads to wave function localization. However,
when an atom in a cloud chamber is ionized by a particle, the exact time of
the event is not measured. Observation of the particle track gives us
approximate positions of the ionized atoms and only indicates that the
particle has passed close to these positions. The particle's wave function
is not necessarily localized in such measurements.

In this paper we are going to argue that the spatial trajectory of a
particle under a potential barrier is just as real as its semi-classical
trajectory in the classically allowed region. The paper is organized as
follows. In the next section we introduce a simple $1$-dimensional model
detector which can perform ``non-destructive'' position measurements. The
detector does not substantially perturb the particle, and yet is able to
register its presence in a particular region. The $3$-dimensional case is
considered in Section \ref{3d-free}. The under-barrier operation of our
detector in one and three dimensions is discussed, respectively, in Sections
\ref{1dbar} and \ref{3d-bar}. The conclusions of the paper are summarized
and discussed in Section \ref{3d-sum}.

\section{Model detector}

The simplest detector is a two-level system with states $|$$0\rangle $ and $%
| $$1\rangle $ which we shall read as ``no particle'' and ``particle
detected'' (or ``up'' and ``down''). In realistic detectors the two states
have different energies, but this energy difference is insignificant as long
as it is much smaller than the other energy scales in the problem. We shall
assume for simplicity that the two states of the detector are degenerate.

The state of the system (which includes a particle and the detector) is
described by a two-component wave function
\begin{equation}
\Psi \left( {\bf r},t\right) =\left[
\begin{array}{c}
u\left(
{\bf r},t\right) \\ d\left( {\bf r},t\right)
\end{array}
\right] ,
\end{equation}
where $u\left( {\bf r},t\right) $ and $d\left( {\bf r},t\right) $ are the
amplitudes for the particle to be at position ${\bf r}$ at time $t$ with the
detector ``up'' and ``down'' respectively.

The interaction between the particle and the detector should flip the state
of the detector when the particle is within the sensitive region. This means
that the interaction introduces non-diagonal elements in the Hamiltonian. We
shall make a simple choice
\begin{equation}
\label{Hint}{\cal H}_{\text{int}}=V\left( {\bf r}\right) \left(
\begin{array}{cc}
0 & 1 \\
1 & 0
\end{array}
\right) \equiv \hat V\left( {\bf r}\right) ,
\end{equation}
where the function $V\left( {\bf r}\right) $ rapidly decreases to zero
outside the sensitivity region.

The total Hamiltonian of the system (we put $\hbar =1$) is given by
\begin{equation}
\label{ham}{\cal H}=\frac{\hat p^2}{2m}+U\left( {\bf r}\right) +{\cal H}_{
\text{int}},
\end{equation}
where $\hat p=-i{\bf \nabla }$ is the momentum operator and $U\left( {\bf r}%
\right) $ is the external potential describing the barrier. We note that the
Hamiltonian (\ref{ham}) is formally equivalent to that of a spin-$1/2$
particle moving in a potential $U\left( {\bf r}\right) $ and in a magnetic
field $B_y=V\left( {\bf r}\right) /q$, $B_x=B_z=0$, where $q$ is the
particle charge. A similar model was employed in \cite{tuntime}, where the
rotation of the particle spin was used to determine the time spent by the
particle under the barrier. For that purpose, the magnetic field in \cite
{tuntime} was chosen to extend over the whole under-barrier region, while
here we are going to assume that the detector size is much smaller than the
barrier width.

We shall first consider the operation of our detector without a barrier,
taking $U\left( {\bf r}\right) =0$. To simplify the discussion here, we
restrict ourselves to the $1$-dimensional case. We will treat the
corresponding $3$-dimensional case in Sections \ref{3d-free} and \ref{3d-bar}%
{}.

To analyze the detection of free particles in one dimension, it is
sufficient to solve the time-independent Schr\"odinger equation
\begin{equation}
\label{tischr}-\frac 1{2m}\frac{d^2\Psi }{dx^2}+\hat V\left( x\right) \Psi
=E\Psi ,
\end{equation}
where $\hat V\left( x\right) $ is taken from (\ref{Hint}). We treat the
detector potential $V\left( x\right) $ as a perturbation, and to linear
order in $V\left( x\right) $ the solution of (\ref{tischr}) is
\begin{equation}
\label{pwf}\Psi \left( x\right) =\Psi ^{\left( 0\right) }\left( x\right)
-\int G\left( x,x^{\prime }\right) \hat V\left( x^{\prime }\right) \Psi
^{\left( 0\right) }\left( x^{\prime }\right) dx^{\prime },
\end{equation}
where
\begin{equation}
G\left( x,x^{\prime }\right) =\frac{im}p\exp \left( ip\left| x-x^{\prime
}\right| \right)
\end{equation}
is the Green's function for the equation (\ref{tischr}), $p\equiv \sqrt{2mE}$%
, and $\Psi ^{\left( 0\right) }\left( x\right) $ is the unperturbed wave
function describing the incident particle with detector ``down'':
\begin{equation}
\label{unpwf}\Psi ^{\left( 0\right) }\left( x\right) =e^{ipx}\left(
\begin{array}{c}
0 \\
1
\end{array}
\right) .
\end{equation}

The asymptotic forms of $\Psi \left( x\right) $ at large $\left| x\right| $
are
\begin{equation}
\label{fasy-p}\Psi \left( x\rightarrow +\infty \right) =e^{ipx}\left( 1-
\frac{im}p\int \hat V\left( x^{\prime }\right) dx^{\prime }\right) \left(
\begin{array}{c}
0 \\
1
\end{array}
\right) ,
\end{equation}
\begin{equation}
\label{fasy-n}\Psi \left( x\rightarrow -\infty \right) =e^{ipx}\left(
1-e^{-2ipx}\frac{im}p\int \hat V\left( x^{\prime }\right) e^{2ipx^{\prime
}}dx^{\prime }\right) \left(
\begin{array}{c}
0 \\
1
\end{array}
\right) .
\end{equation}
The expressions (\ref{fasy-p}), (\ref{fasy-n}) directly give us the
probabilities of different interaction outcomes. The term multiplied by the
``up'' detector state describes the wave function of the particle in case it
is detected, and hence the probability of transmission with detection is
\begin{equation}
\label{w-td}w_{\text{td}}=\left| \frac mp\int V\left( x\right) dx\right| ^2,
\end{equation}
and the probability of reflection with detection is
\begin{equation}
\label{w-rd}w_{\text{rd}}=\left| \frac mp\int V\left( x\right)
e^{2ipx}dx\right| ^2 .
\end{equation}
(There is no reflection without detection in the first-order approximation.)
If the size $L$ of the interaction region is large enough so that $L\gg
p^{-1}$, then the integral in (\ref{w-rd}) is much smaller than the
corresponding integral in (\ref{w-td}). This means that $w_{\text{r}}\ll w_{
\text{td}}$ and hence the total probability of detection is
\begin{equation}
\label{wd}w_{\text{d}}=w_{\text{td}}+w_{\text{rd}}\approx w_{\text{td}%
}=\left| \frac mp\int V\left( x\right) dx\right| ^2.
\end{equation}

The detection probability (\ref{wd}) may be written as
\begin{equation}
\label{rwd}w_{\text{d}}=\left( \bar V\Delta t\right) ^2,
\end{equation}
where
\begin{equation}
\label{rav}\bar V=\frac 1L\int V\left( x\right) dx
\end{equation}
is the average interaction potential and
\begin{equation}
\label{rdt}\Delta t=\frac{mL}p
\end{equation}
is the time it takes the particle to traverse the detector location (the
interaction region of size $L$). We will later compare this expression (\ref
{rwd}) with the under-barrier case in Section \ref{1dbar}.

The detection efficiency can be improved by increasing the number of
internal states of the detector. A simple extension of our model (\ref{Hint}%
), (\ref{ham}) with a detector having $N$ internal states is constructed by
taking the interaction term $\hat V\left( x\right) =V\left( x\right) \hat M$%
, where the matrix $\hat M$ must have non-zero off-diagonal elements. For
instance, if we choose for simplicity $M_{ij}=1-\delta _{ij}$,
\begin{equation}
\hat V\left( x\right) =V\left( x\right) \left(
\begin{array}{cccc}
0 & 1 & \cdots & 1 \\
1 & 0 & \cdots & \vdots \\
\vdots & \vdots & \ddots & 1 \\
1 & \cdots & 1 & 0
\end{array}
\right)
\end{equation}
the detection probability (\ref{wd}) will be multiplied by $\left(
N-1\right) $. With a suitable choice of parameters the detection probability
could be made close to $1$, while the reflection probability would still be
negligible. (The perturbative expressions like (\ref{w-rd}), (\ref{wd}) are
valid only when the detection probability is very small, $w_{\text{d}}\ll 1$%
; nevertheless one can show using exactly solvable potentials that it is
indeed possible to make $w_{\text{d}}$ close to $1$.)

\section{\label{1dbar}Under-barrier detection in 1 dimension}

In order to introduce some useful relations we shall first consider barrier
penetration without a detector. The corresponding Schr\"odinger equation in
one dimension is
\begin{equation}
\label{tbschr}-\frac 1{2m}\frac{d^2\psi }{dx^2}+U\left( x\right) \psi =E\psi
{}.
\end{equation}
For a particle incident on the barrier in the positive $x$ direction, the
asymptotic forms of the wave function $\psi ^{+}\left( x\right) $ for large $%
\left| x\right| $ are:
\begin{equation}
\label{asy-p}
\begin{array}{c}
\psi ^{+}\left( x\rightarrow -\infty \right) =e^{ipx}+Be^{-ipx}, \\
\psi ^{+}\left( x\rightarrow \infty \right) =Ce^{ipx}.
\end{array}
\end{equation}
The tunneling probability is $w_{\text{t}}=\left| C\right| ^2$. For a
particle incident in the negative $x$ direction, the wave function $\psi
^{-}\left( x\right) $ is
\begin{equation}
\label{asy-n}
\begin{array}{c}
\psi ^{-}\left( x\rightarrow -\infty \right) =C^{\prime }e^{-ipx}, \\
\psi ^{-}\left( x\rightarrow \infty \right) =e^{-ipx}+B^{\prime }e^{ipx}.
\end{array}
\end{equation}
and it can be shown \cite{first} that $\left| B^{\prime }\right| =\left|
B\right| $, $\left| C^{\prime }\right| =\left| C\right| $. We shall assume
for simplicity that the barrier is symmetric, $U\left( -x\right) =U\left(
x\right) $. Then $C^{\prime }=C$, $B^{\prime }=B$, and
\begin{equation}
\label{ppm}\psi ^{-}\left( x\right) =\psi ^{+}\left( -x\right) .
\end{equation}

An approximate form of the tunneling wave functions $\psi ^{\pm }\left(
x\right) $ can be found in the WKB limit. Let the classical turning points
of the potential be $x=b$ and $x=-b$, so $U\left( \pm b\right) =E$. Then
\cite{first}
\begin{equation}
\psi ^{+}\left( x>b\right) \approx C\sqrt{\frac p{k\left( x\right) }}\exp
\left[ i\int\limits_b^xk\left( x^{\prime }\right) dx^{\prime }+\frac{i\pi }%
4\right] ,
\end{equation}
\begin{equation}
\label{pib}\psi ^{+}\left( -b<x<b\right) \approx C\sqrt{\frac p{k\left(
x\right) }}\exp \left[ \int\limits_x^bk\left( x^{\prime }\right) dx^{\prime
}\right] ,
\end{equation}
where
\begin{equation}
k\left( x\right) =\sqrt{2m\left| E-U\left( x\right) \right| }.
\end{equation}
A similar expression can be written for $x<-b$. The tunneling amplitude in
the WKB approximation is given by
\begin{equation}
\label{ta}C=\exp \left[ -\int\limits_{-b}^bk\left( x\right) dx\right] .
\end{equation}

The wave function $\psi ^{-}\left( x\right) $ for a particle incident in the
opposite direction can be found from (\ref{ppm}). It is then easy to show
that for $x$ inside the barrier, $-b<x<b$, the functions $\psi ^{\pm }\left(
x\right) $ satisfy the relation
\begin{equation}
\label{pppm}\psi ^{+}\left( x\right) \psi ^{-}\left( x\right) =C\frac
p{k\left( x\right) },\quad -b<x<b.
\end{equation}

In the presence of a detector, the Schr\"odinger equation becomes
\begin{equation}
\label{tbdschr}\left( -\frac 1{2m}\frac{d^2}{dx^2}+U\left( x\right) +\hat
V\left( x\right) \right) \Psi =E\Psi .
\end{equation}
To linear order in $V\left( x\right) $, its solution can still be written in
the form (\ref{pwf}), where now
\begin{equation}
\label{upwf}\Psi ^{\left( 0\right) }\left( x\right) =\psi ^{+}\left(
x\right) \left(
\begin{array}{c}
0 \\
1
\end{array}
\right)
\end{equation}
and $G\left( x,x^{\prime }\right) $ is a Green's function of the equation (%
\ref{tbschr}):
\begin{equation}
\label{gfeq}\left( -\frac 1{2m}\frac{d^2}{dx^2}+U\left( x\right) -E\right)
G\left( x,x^{\prime }\right) =\delta \left( x-x^{\prime }\right) .
\end{equation}

The Green's function $G\left( x,x^{\prime }\right) $ is determined by eq. (%
\ref{gfeq}) only up to an arbitrary solution of the homogeneous equation (%
\ref{tbschr}). This freedom can be removed by imposing suitable boundary
conditions at $x\rightarrow \pm \infty $. The boundary conditions
appropriate for a scattering problem require that $G\left( x,x^{\prime
}\right) $ at $x\rightarrow \pm \infty $ contains only outgoing waves, that
is only terms proportional to $e^{ipx}$ at $x\rightarrow +\infty $ and to $%
e^{-ipx}$ at $x\rightarrow -\infty $. With this choice of boundary
conditions, the Green's function can be expressed in terms of the tunneling
solutions $\psi ^{\pm }\left( x\right) $,
\begin{equation}
\label{gf}G\left( x,x^{\prime }\right) =-\frac{2m}W\cdot \left\{
\begin{array}{l}
\psi ^{+}\left( x\right) \psi ^{-}\left( x^{\prime }\right) ,\quad x\geq
x^{\prime }, \\
\psi ^{-}\left( x\right) \psi ^{+}\left( x^{\prime }\right) ,\quad x\leq
x^{\prime }.
\end{array}
\right.
\end{equation}
Here $W$ is the Wronskian which can be evaluated using the asymptotic
expressions (\ref{asy-p}), (\ref{asy-n}):
\begin{equation}
\label{wron}W=\psi ^{+\prime }\left( x\right) \psi ^{-}\left( x\right) -\psi
^{+}\left( x\right) \psi ^{-\prime }\left( x\right) =2ipC.
\end{equation}
Combining eqs. (\ref{pwf}), (\ref{upwf}), (\ref{gf}), and (\ref{wron}) we
can find the asymptotic forms of the wave function of the system to the
first order in $V\left( x\right) $:
\begin{equation}
\label{asyb-p}\Psi \left( x\rightarrow +\infty \right) =e^{ipx}\left[
C\left(
\begin{array}{c}
0 \\
1
\end{array}
\right) -\frac{im}p\left(
\begin{array}{c}
1 \\
0
\end{array}
\right) \int V\left( x^{\prime }\right) \psi ^{+}\left( x^{\prime }\right)
\psi ^{-}\left( x^{\prime }\right) dx^{\prime }\right] ,
\end{equation}
\begin{equation}
\label{asyb-n}\Psi \left( x\rightarrow -\infty \right) =e^{ipx}\left(
\begin{array}{c}
0 \\
1
\end{array}
\right) +e^{-ipx}\left[ B\left(
\begin{array}{c}
0 \\
1
\end{array}
\right) -\frac{im}p\left(
\begin{array}{c}
1 \\
0
\end{array}
\right) \int V\left( x^{\prime }\right) {\psi ^{+}}^2\left( x^{\prime
}\right) dx^{\prime }\right] .
\end{equation}

For each incident particle there are four possible outcomes: the particle is
either transmitted or reflected and it is either detected or not. The
probabilities of these events which we denote $w_{\text{dt}}$, $w_{\text{dr}%
} $, $w_{\text{nt}}$, and $w_{\text{nr}}$, can be found directly from eqs. (%
\ref{asyb-p})--(\ref{asyb-n}). We shall be interested in conditional
probabilities of detection for transmitted and reflected particles, $w\left(
\text{d}|\text{t}\right) $ and $w\left( \text{d}|\text{r}\right) $. For
instance, $w\left( \text{d}|\text{t}\right) $ is the probability of
detection under the condition that the particle is transmitted; it is given
by
\begin{equation}
w\left( \text{d}|\text{t}\right) =\frac{w_{\text{dt}}}{w_{\text{t}}},
\end{equation}
where $w_{\text{t}}=w_{\text{dt}}+w_{\text{nt}}$ is the total transmission
probability.

We now shall explore the dependence of these probabilities on the position
of the detector under the barrier. We assume that the size of the sensitive
region is much smaller than the barrier width, $L\ll b$.

{}From (\ref{asyb-n}), the detection probability for a reflected particle is
\begin{equation}
\label{wbdr}w\left( d|r\right) =\left| \frac mp\int V\left( x\right) \psi
^{+2}\left( x\right) dx\right| ^2,
\end{equation}
where we have assumed that the tunneling probability $\left| C\right| ^2$ is
small so that we may take $\left| B\right| ^2\approx 1$. We see that the
probability (\ref{wbdr}) is highest when the detector is placed near the
classical turning point $x=-b$, and that $w\left( d|r\right) $ decreases
exponentially as the detector is moved further under the barrier:
\begin{equation}
w\left( d|r\right) \sim \left| \psi ^{+}\left( x_0\right) \right| ^4,
\end{equation}
where $x_0$ is the detector location.

The detection probability for transmitted particles, $w\left( d|t\right) $,
can be obtained from (\ref{asyb-p}):
\begin{equation}
\label{wbdt}w\left( d|t\right) =\left| m\int \frac{V\left( x\right) }{%
k\left( x\right) }dx\right| ,
\end{equation}
where we have used the WKB relation (\ref{pppm}). As in the previous
section, we can write [cf. eqs. (\ref{rwd})--(\ref{rdt})]
\begin{equation}
\label{ewd}w\left( d|t\right) \approx \left( \frac{mL}k\bar V\right)
^2=\left( \bar V\Delta t\right) ^2,
\end{equation}
where $k\left( x\right) $ is calculated at the detector location and $\Delta
t=mL/k$ can be interpreted as the Euclidean detector traversal time, i.e.
the time it takes to traverse a length $L\,$ with the Euclidean momentum $k$.

Note that unlike the reflection probability $w\left( d|r\right) $, the
detection probability (\ref{wbdt}) is not exponentially suppressed under the
barrier and depends on the detector position only through $k\left( x\right) $%
. [It should be kept in mind that the WKB-based relation (\ref{pppm}) which
gave rise to the simple expression (\ref{wbdt}) is not valid near the
classical turning points $x=\pm b$.]

An intuitive explanation of these results can be suggested. The reflection
amplitude can be obtained in the path integral approach by summing over
paths that originate at and return back to the negative infinity. This
amplitude is dominated by the vicinity of a classical path in which the
particle is reflected from the barrier at $x=-b$. A detector placed far
under the barrier is not traversed by such paths, so the detection
probability is exponentially suppressed. On the other hand, the transmission
amplitude is obtained by summation over paths originating at $x=-\infty $
and ending at $x=\infty $. All such paths traverse the location of the
detector, and so the detection probability is high regardless of the
detector position.

Suppose now that a large number of detectors is uniformly distributed in the
classically forbidden region $-b<x<b$ and that a particle is incident on the
barrier from the negative $x$ direction. After interaction some of the
detectors will register the particle; these excited detectors form the
``track'' which we observe. Then if the particle is reflected, we expect
that only some of the detectors near $x=-b$ will register the particle. On
the other hand, if the particle is transmitted, the track will stretch
throughout the entire under-barrier region, with the highest density of
detection near the turning points $x=\pm b$ where the Euclidean momentum $%
k\left( x\right) $ is small.

\section{\label{3d-free}Free particle detection in 3 dimensions}

We now consider detection of a free particle in $3$ dimensions. The
Schr\"odinger equation for the detector-particle system is
\begin{equation}
\label{3dschr}\left( -\frac 1{2m}{\bf \nabla }^2+\hat V\left( {\bf r}\right)
\right) \Psi \left( {\bf r}\right) =E\Psi \left( {\bf r}\right) .
\end{equation}

We assume as before that the operator $\hat V\left( {\bf r}\right) $ is of
the form (\ref{Hint}) and that the detector potential $V\left( {\bf r}%
\right) $ rapidly decreases outside a region of size $L$. The coordinates
can always be chosen so that the detector is centered at the origin and the
incoming particle moves in the positive $x$ direction. As an example of the
detector potential function we can choose
\begin{equation}
\label{gaup}V\left( {\bf r}\right) =V_0\exp \left( -\frac{r^2}{L^2}\right) ,
\end{equation}
but in most of the following discussion we will not have to assume a
specific form of $V\left( {\bf r}\right) $.

To linear order in $V\left( {\bf r}\right) $, the solution of (\ref{3dschr})
is given by
\begin{equation}
\label{3dpwf}\Psi \left( {\bf r}\right) =\Psi ^{\left( 0\right) }\left( {\bf %
r}\right) -\int G\left( {\bf r},{\bf r}^{\prime }\right) \hat V\left( {\bf r}%
^{\prime }\right) \Psi ^{\left( 0\right) }\left( {\bf r}^{\prime }\right) d^3%
{\bf r}^{\prime },
\end{equation}
where the unperturbed wave function $\Psi ^{\left( 0\right) }\left( {\bf r}%
\right) $ is still given by (\ref{unpwf}),
\begin{equation}
\Psi ^{\left( 0\right) }\left( {\bf r}\right) =e^{ipx}\left(
\begin{array}{c}
0 \\
1
\end{array}
\right) ,
\end{equation}
and the Green's function $G\left( {\bf r},{\bf r}^{\prime }\right) $
satisfies the equation
\begin{equation}
\label{3dgfeq}\left( -\frac 1{2m}{\bf \nabla }^2-E\right) G\left( {\bf r},%
{\bf r}^{\prime }\right) =\delta \left( {\bf r}-{\bf r}^{\prime }\right) .
\end{equation}
The solution of (\ref{3dgfeq}) with the ``outgoing wave'' boundary condition
at infinity is
\begin{equation}
G\left( {\bf r},{\bf r}^{\prime }\right) =\frac 1{4\pi \left| {\bf r}-{\bf r}%
^{\prime }\right| }\exp \left( ip\left| {\bf r}-{\bf r}^{\prime }\right|
\right)
\end{equation}
with $p=\sqrt{2mE}$.

At large distances from the detector ($r\gg L$), the integration over ${\bf r%
}^{\prime }$ in (\ref{3dpwf}) is effectively performed over the range $%
r^{\prime }\ll r$, and we can use the expansion
\begin{equation}
\left| {\bf r}-{\bf r}^{\prime }\right| =r-{\bf n}\cdot {\bf r}^{\prime
}+O\left( r^{\prime 2}\right) ,
\end{equation}
where ${\bf n}$ is a unit vector in the direction of ${\bf r}$. This gives
\begin{equation}
\label{3dapwf}\Psi \left( {\bf r}\right) \approx \left(
\begin{array}{c}
0 \\
1
\end{array}
\right) e^{ipx}-\left(
\begin{array}{c}
1 \\
0
\end{array}
\right) \frac{e^{ipr}}{4\pi r}\int e^{ip\left( x^{\prime }-{\bf n}\cdot {\bf %
r}^{\prime }\right) }V\left( {\bf r}^{\prime }\right) d^3{\bf r}^{\prime }.
\end{equation}

We shall assume that the detector is large compared to the particle
wavelength, that is
\begin{equation}
L\gg p^{-1}.
\end{equation}
Then the ${\bf r}^{\prime }$-integral in (\ref{3dapwf}) is suppressed by the
rapidly oscillating exponential, unless ${\bf n}$ is nearly parallel to the $%
x$-axis, so that the angle $\theta $ between ${\bf n}$ and the $x$ axis
satisfies
\begin{equation}
\left| pL\sin \theta \right| \lesim 1.
\end{equation}
As a result, the particle wave function corresponding to the excited
detector in (\ref{3dapwf}) is appreciably non-zero only in a narrow cone of
angular width $\theta _{\max }\sim \left( pL\right) ^{-1}$ as shown on Fig.~$%
1$.

The interpretation of this result is straightforward. The interaction with
the detector localizes the particle within a region of size $L$, resulting
in a transverse momentum uncertainty $p_{\perp }\sim L^{-1}$. Therefore the
direction of motion after interaction may be altered by an angle
\begin{equation}
\theta _{\max }\sim p_{\perp }/p\sim \left( pL\right) ^{-1}.
\end{equation}

If we distribute detectors uniformly in space, then we expect the moving
particle to create a track in which (almost) every excited detector lies
within a cone of angular width $\theta _{\max }$ from the preceding excited
detector. This can be verified directly \cite{second} by calculating the
probability for the particle to be registered by two detectors, using the
second-order perturbation expansion in $V\left( {\bf r}\right) $.

\section{\label{3d-bar}Under-barrier detection in 3 dimensions}

Turning now to the $3$-dimensional barrier penetration problem, we assume
the geometry illustrated in Fig. $2$. The particle is described by a plane
wave traveling in the positive $x$-direction and the barrier potential $%
U\left( {\bf r}\right) $ is a function only of $x$. Hence, without the
detector the problem reduces to the one-dimensional problem of Section \ref
{1dbar}. We shall assume that the detector is centered on the $x$-axis at $%
x=x_0$, with $-b<x_0<b$, and that its size is much smaller than the barrier
width, $L\ll b$.

The solution of the Schr\"odinger equation
\begin{equation}
\label{3dbschr}\left( -\frac 1{2m}{\bf \nabla }^2+U\left( x\right) +\hat
V\left( {\bf r}\right) \right) \Psi \left( {\bf r}\right) =E\Psi \left( {\bf %
r}\right)
\end{equation}
can still be written in the form (\ref{3dpwf}), where now the Green's
function satisfies the equation
\begin{equation}
\label{3dgfbeq}\left( -\frac 1{2m}{\bf \nabla }^2+U\left( x\right) -E\right)
G\left( {\bf r},{\bf r}^{\prime }\right) =\delta \left( {\bf r}-{\bf r}%
^{\prime }\right)
\end{equation}
and
\begin{equation}
\Psi ^{\left( 0\right) }=\psi ^{+}\left( x\right) \left(
\begin{array}{c}
0 \\
1
\end{array}
\right) .
\end{equation}
The symmetry of the problem suggests that the dependence of $G\left( {\bf r},%
{\bf r}^{\prime }\right) $ on $y,y^{\prime }$ and $z,z^{\prime }\,$ is
reduced to the dependence on the distance in the $\left( y,z\right) $-plane.
Hence, we can represent it as a Fourier transform,
\begin{equation}
G\left( {\bf r},{\bf r}^{\prime }\right) =\int \frac{d^2{\bf q}}{\left( 2\pi
\right) ^2}G_q\left( x,x^{\prime }\right) e^{i{\bf q}\left( {\bf R}-{\bf R}%
^{\prime }\right) }.
\end{equation}
Here, ${\bf R}\equiv \left( y,z\right) $, ${\bf R}^{\prime }\equiv \left(
y^{\prime },z^{\prime }\right) $ and $G_q\left( x,x^{\prime }\right) $
satisfies
\begin{equation}
\label{3dgfq}\left( -\frac 1{2m}\frac{d^2}{dx^2}+U\left( x\right)
-E_q\right) G_q\left( x,x^{\prime }\right) =\delta \left( x-x^{\prime
}\right) ,
\end{equation}
where
\begin{equation}
\label{en-q}E_q=E-\frac{q^2}{2m}.
\end{equation}

Equation (\ref{3dgfq}) is identical to eq. (\ref{gfeq}) for the $1$%
-dimensional problem, and we can read the solution from (\ref{gf}), (\ref
{wron}):
\begin{equation}
\label{qgf}G_q\left( x,x^{\prime }\right) =\frac{im}{p_qC_q}\cdot \left\{
\begin{array}{l}
\psi _q^{+}\left( x\right) \psi _q^{-}\left( x^{\prime }\right) ,\quad x\geq
x^{\prime }, \\
\psi _q^{-}\left( x\right) \psi _q^{+}\left( x^{\prime }\right) ,\quad x\leq
x^{\prime }.
\end{array}
\right.
\end{equation}
Here, the subscripts $q$ indicate that the corresponding quantities are
evaluated for the energy $E_q$ given by (\ref{en-q}).

Suppose the detector is centered at $x=x_0$, ${\bf R}=0$. Then for $x-x_0\gg
L$ we can use the upper line of (\ref{qgf}) for the Green's function in eq. (%
\ref{3dpwf}). Using the WKB expressions (\ref{pib}) for $\psi _q^{+}\left(
x\right) $ and $\psi _q^{-}\left( x\right) =\psi _q^{+}\left( -x\right) $
together with eqs. (\ref{ppm}), (\ref{ta}), we obtain after some algebra:%
$$
\Psi \left( {\bf r}\right) =\psi ^{+}\left( x\right) \left(
\begin{array}{c}
0 \\
1
\end{array}
\right) -\left(
\begin{array}{c}
1 \\
0
\end{array}
\right) im\int \frac{d^2{\bf q}}{\left( 2\pi \right) ^2}\frac{C_0\sqrt{p_0}}{%
C_q\sqrt{p_q}}e^{i{\bf qR}}\psi _q^{+}\left( x\right)
\TeXButton{horizontal-skip}{\hskip 1.5 in}
$$
\begin{equation}
\label{wf1}\TeXButton{horizontal-skip}{\hskip 2 in}\cdot \int \frac{%
dx^{\prime }V_q\left( x^{\prime }\right) }{\sqrt{k_q\left( x^{\prime
}\right) k_0\left( x^{\prime }\right) }}\exp \left\{ -\int\limits_{x^{\prime
}}^b\left[ k_q\left( \xi \right) -k_0\left( \xi \right) \right] d\xi
\right\} ,
\end{equation}
where
\begin{equation}
V_q\left( x\right) =\int d^2{\bf R}e^{-i{\bf qR}}V\left( x,{\bf R}\right)
\end{equation}
and a subscript ``$0$'' indicates that the corresponding quantities are
taken at ${\bf q}=0$.

The Fourier component $V_q\left( x\right) $ is small for $q\gg L^{-1}$. For
instance, with the detector potential (\ref{gaup}),
\begin{equation}
V_q\left( x\right) =\pi L^2V_0\exp \left( -\frac{q^2L^2}4\right) \exp \left(
-\frac{\left( x-x_0\right) ^2}{L^2}\right) .
\end{equation}
We shall assume that $L^{-1}\ll k_0\left( x\right) $ almost everywhere under
the barrier, except in the vicinity of the turning points $x=\pm b$. If we
don't place the detector near $x=\pm b$ (which is necessary also because the
WKB approximation breaks down at those points) then the integration in (\ref
{wf1}) is over $x^{\prime }$ which are far from $x^{\prime }=\pm b$. For
these $x^{\prime }$ we can put $k_q\left( x^{\prime }\right) \approx
k_0\left( x^{\prime }\right) $ and replace $k_q$ by $k_0$ everywhere in (\ref
{wf1}) except in the exponential, where $k_q$ can be expanded as
\begin{equation}
\label{kqk0}k_q\approx k_0+\frac{q^2}{2k_0}.
\end{equation}
The expansion (\ref{kqk0}) is not valid all the way up to the turning point,
but one can show that (\ref{kqk0}) gives a good approximation for the $\xi $%
-integral in (\ref{wf1}) if $q$ is much smaller than typical values of $k_0$%
. Since $p_q=\sqrt{2mE-q^2}$, for $q\ll \sqrt{2mE}$ we can also set $%
p_q\approx p_0$ (although we cannot substitute ${\bf q}=0$ in $C_q$ because
the dependence of the latter on $q$ is exponential). The resulting
expression is
\begin{equation}
\label{se13}\Psi \left( {\bf r}\right) =\psi ^{+}\left( x\right) \left(
\begin{array}{c}
0 \\
1
\end{array}
\right) -\left(
\begin{array}{c}
1 \\
0
\end{array}
\right) im\int \frac{d^2{\bf q}}{\left( 2\pi \right) ^2}\frac{C_0}{C_q}e^{i%
{\bf qR}}\psi _q^{+}\left( x\right) \cdot \int \frac{dx^{\prime }V_q\left(
x^{\prime }\right) }{k_0\left( x^{\prime }\right) }\exp \left\{ -\frac{q^2}%
2\int\limits_{x^{\prime }}^b\frac{d\xi }{k_0\left( \xi \right) }\right\} .
\end{equation}

To find the large $x$ asymptotics of the wave function, we replace $\psi
_q^{+}\left( x\right) $ in (\ref{se13}) by its asymptotic form (\ref{asy-p}%
),
\begin{equation}
\psi _q^{+}\left( x\right) =C_qe^{ip_qx}.
\end{equation}
For very large $x$ and ${\bf R}$, the ${\bf q}$-integration is dominated by
the stationary point of the phase, ${\bf qR}+p_qx$, which is given by
\begin{equation}
{\bf Q}=\frac{{\bf R}}{\sqrt{R^2+x^2}}p_0={\bf n}p_0\sin \theta ,
\end{equation}
where ${\bf n}$ is a unit vector pointing in the direction of ${\bf R}$ and $%
\theta $ is the angle between ${\bf r}\equiv \left( x,{\bf R}\right) $ and
the $x$ axis, $\tan \theta =R/x.$ Using the stationary phase approximation,
we find for $x\rightarrow +\infty $:
\begin{equation}
\label{415}\Psi \left( {\bf r}\right) =\left(
\begin{array}{c}
0 \\
1
\end{array}
\right) Ce^{ipx}-\left(
\begin{array}{c}
1 \\
0
\end{array}
\right) \frac{e^{ipr}\cos \theta }{2\pi r}Cmp\int \frac{dx^{\prime
}V_Q\left( x^{\prime }\right) }{k\left( x^{\prime }\right) }\exp \left[ -
\frac{Q^2}2\int\limits_{x^{\prime }}^b\frac{d\xi }{k\left( \xi \right) }%
\right] ,
\end{equation}
where the subscript ``$0$'' has been dropped.

The integral over $x^{\prime }$ in (\ref{415}) can be estimated as
\begin{equation}
\label{iq}I_Q\left( x_0,b\right) \sim \frac{\bar V_QL}{k\left( x_0\right) }%
\exp \left[ -\frac{Q^2}2\int\limits_{x_0}^b\frac{d\xi }{k\left( \xi \right) }%
\right] ,
\end{equation}
where $\bar V_Q$ is defined as in (\ref{rav}). Large values of $Q$ are
suppressed in $I_Q$ both by $\bar V_Q$ and by the exponential factor, so we
get a bell-shaped function of $Q$. As a result, the particle wave function
corresponding to the excited detector state $\left(
\begin{array}{c}
1 \\
0
\end{array}
\right) $ in (\ref{415}) describes a narrow beam of angular width
\begin{equation}
\theta \approx \frac{Q_{\max }}p,
\end{equation}
where $Q_{\max }$ is the effective cutoff introduced by $I_Q$. If the
detector is close to the end of the barrier at $x=b$, so that
\begin{equation}
\label{width}\int\limits_{x_0}^b\frac{d\xi }{k\left( \xi \right) }\equiv
\left( b-x_0\right) \overline{k^{-1}\left( x_0,b\right) }\lesim L^2,
\end{equation}
then
\begin{equation}
Q_{\max }\sim L^{-1}.
\end{equation}
Otherwise,
\begin{equation}
Q_{\max }\sim \left[ \left( b-x_0\right) \overline{k^{-1}\left( x_0,b\right)
}\right] ^{-\frac 12}.
\end{equation}

The under-barrier wave function in the range $x_0<x<b$ can be obtained from (%
\ref{se13}) by substituting the WKB expression (\ref{pib}) for $\psi
_q^{+}\left( x\right) $. This gives
\begin{equation}
\label{421}\Psi \left( {\bf r}\right) =\left(
\begin{array}{c}
0 \\
1
\end{array}
\right) \psi ^{+}\left( x\right) -\left(
\begin{array}{c}
1 \\
0
\end{array}
\right) \psi ^{+}\left( x\right) im\int \frac{d^2{\bf q}}{\left( 2\pi
\right) ^2}e^{i{\bf qR}}\int \frac{dx^{\prime }V_q\left( x^{\prime }\right)
}{k\left( x^{\prime }\right) }\exp \left[ -\frac{q^2}2\int\limits_{x^{\prime
}}^x\frac{d\xi }{k\left( \xi \right) }\right] .
\end{equation}
Just like in (\ref{iq}), the integral over $x^{\prime }$ in (\ref{421}) is a
bell-shaped function of ${\bf q}$, $I_q\left( x_0,x\right) $, but now it has
an $x$-dependent width expressed in the notation of (\ref{width}) as
\begin{equation}
q_{\max }\left( x\right) \sim \min \left\{ L^{-1},\left[ \left( x-x_0\right)
\overline{k^{-1}\left( x_0,x\right) }\right] ^{-\frac 12}\right\} .
\end{equation}
This width, in turn, determines the spatial width of the region in which the
wave function of the particle with an excited detector is appreciably
different from zero:
\begin{equation}
R_{\max }\left( x\right) \sim q_{\max }^{-1}\left( x\right) =\max \left\{
L,\left[ \left( x-x_0\right) \overline{k^{-1}\left( x_0,x\right) }\right]
^{\frac 12}\right\} .
\end{equation}
Beyond the barrier, at $x>b$, the width $R_{\max }$ is growing linearly in $x
$:
\begin{equation}
R_{\max }\left( x>b\right) =R_{\max }\left( b\right) +\left( x-b\right)
\theta _{\max },
\end{equation}
where $\theta _{\max }=q_{\max }\left( b\right) /p$ as in Section~\ref
{3d-free}.

Using a similar argument, one finds that the wave function of the detected
particle in the range $-b<x<x_0$ is exponentially damped compared to the
unperturbed wave function $\psi ^{+}\left( x\right) $:
\begin{equation}
\Psi \left( {\bf r}\right) =\left(
\begin{array}{c}
0 \\
1
\end{array}
\right) \psi ^{+}\left( x\right) -\left(
\begin{array}{c}
1 \\
0
\end{array}
\right) \psi ^{+}\left( x\right) \exp \left( -2\int\limits_x^{x_0}k\left(
\xi \right) d\xi \right) im\int \frac{d^2{\bf q}}{\left( 2\pi \right) ^2}e^{i%
{\bf qR}}I_q\left( x,x_0\right) .
\end{equation}
The region where the wave function of the particle with an excited detector
is not negligibly small is outlined in Fig. $3$. The wave function is
exponentially decreasing with the characteristic length $k^{-1}$ in both
directions along the $x$ axis (and with $x$-dependent characteristic length
$R_{\max }\left( x\right) $ in all transverse directions) and has its
maximum at the detector location. Note, however, that the exponential decay
of the wave function in the $x$ direction does not affect the conditional
probability of detection for transmitted particles (since the unperturbed
wave function without a detector decays in the same manner).

To observe the track of a tunneling particle, we distribute detectors
uniformly in the under-barrier region. If the particle has tunneled out of
the barrier and was registered at a position $(x,y,z)$, where $x>b$, then
the conditional probability of a detector registering the particle under the
barrier at a position $(x_0,y_0,z_0)$ will be
vanishing unless $(y,z)$ lies within the distance $R_{\max }\left( x\right) $
of $(y_0,z_0)$. Therefore, the detectors registering the
particle under the barrier will form a line going through the barrier
approximately parallel to the $x$ axis, and the track of the particle
outside the barrier will be a continuation of this line.

If $N$ detectors register the tunneling particle, the deviation of the
particle in the transverse direction will be approximately $\sqrt{N}\cdot
R_{\max }\left( \Delta x\right) $, where $\Delta x$ is the average distance
between two excited detectors. We can compare this deviation to that
obtained in the free case in Section~\ref{3d-free}. Assuming that the
average Euclidean momentum $k$ under barrier is of the same order as the
particle momentum $p$ in the free case, one can show that the under-barrier
track is always narrower than the free particle track.

In actual measurements, the track width will probably be determined by the
bubble size in a bubble chamber or by the photo-emulsion grain size rather
than by transverse deviations from the $x$ direction.

\section{\label{3d-sum}Discussion}

We have considered a model detector which registers semiclassical particles
traversing its location without substantially perturbing the energy of the
particles. If placed sufficiently far under a potential barrier, such
detectors will be sensitive to tunneling particles but will not detect
reflected particles.

By distributing many detectors in a region of space, we can observe the
particle's track formed by the detectors registering the particle. As
expected, the free particle track obtained in this manner is a straight
line. In the presence of a potential barrier, the reflected particles create
a track which does not extend much beyond the classical turning point,
whereas tunneling particles will create a continuous track going through the
barrier region. In the case of a plane-symmetric barrier with a particle
moving orthogonally to the plane, the track is a nearly straight line. In a
more general setting we expect the track to follow the Euclidean trajectory
of the particle.

Having established that the under-barrier path of a particle is a
(semiclassically) meaningful concept, one can now ask whether or not any
meaning can be assigned to a spacetime trajectory in the under-barrier
region. As we emphasized in the Introduction, the readings of our detectors
give no information about the times when the particle passed the detector
locations. One could try to determine these times by using time-dependent
detectors which are turned on for a finite time interval $\Delta \tau $ and
then turned off. For very small $\Delta $$\tau $, however, the perturbation
of the particle energy incurred by the detector (of order $\Delta $$\tau
^{-1}$) would throw the particle above the potential barrier and destroy
tunneling. Therefore, we cannot time the tunneling process more precisely
than up to $\Delta $$\tau _{\min }\sim $$\left( V_0-E\right) ^{-1}$. This
brings about the somewhat controversial issue of the tunneling time (see for
instance \cite{tuntimes} and references therein). Several definitions of the
tunneling time have been suggested \cite{tuntime}, but the most relevant for
us here is the {\em dwell time} $\tau _D$, which is defined as the average
time spent by the particle under the barrier. In the nonrelativistic case
this time is essentially independent of the barrier width, and for a
rectangular barrier is given by \cite{tuntimes}
\begin{equation}
\tau _D=\frac 1{V_0}\sqrt{\frac E{V_0-E}}.
\end{equation}
It is easily verified that $\tau _D$ is always smaller than the measurement
limit $\Delta $$\tau _{\min }$, and thus non-destructive timing of the
tunneling particle is impossible.

In the relativistic case, however, causality requires that a barrier of
width $2b$ cannot be traversed in a time less than $2b/c$. For a
sufficiently wide barrier, this time can be arbitrarily large, suggesting
that the measurement limit $\Delta $$\tau _{\min }$ can be circumvented. If
relativistic under-barrier motion can indeed be timed, it would be
interesting to use time-dependent detectors to analyze vacuum decay
processes such as pair production in an electric field or bubble nucleation
in a false vacuum. In these two cases, the evolution after tunneling is
Lorentz-invariant, and appears the same to all inertial observers. It is not
clear whether or not this applies to the under-barrier evolution as well.
Detectors introduce a preferred reference frame, and it is quite possible
that the evolution seen by an observer will depend on her velocity with
respect to that frame.

Another problem that can be studied using model detectors similar to ours is
the problem of under-barrier measurement in quantum cosmology. Tunneling of
the whole Universe has been extensively discussed in the literature \cite
{cosmology}, and it has been conjectured that the origin of the Universe was
a quantum tunneling event. The tunneling time controversy does not exist in
quantum cosmology: there are no clocks external to the Universe, and the
cosmological wave function does not depend on ``time''. (In fact, ``time''
is an arbitrary coordinate label in general relativity, and physics should
not depend on it.) A physically meaningful time is defined using the
geometric or matter degrees of freedom (e.g., the radius of the Universe,
the abundance of radioactive elements, or detectors like ours). It would be
interesting to investigate how under-barrier evolution will be seen by
internal observers (or how it will be registered by internal detectors).

\medskip\bigskip {\Large {\bf Acknowledgements}} \bigskip

We would like to thank Slava Mukhanov for stimulating our interest in this
problem and for useful discussions in the course of our research. We also
gratefully acknowledge discussions with Francis Low and Richard Milburn.
A.V. is grateful to Alan Guth for his hospitality at MIT, where this research
was completed.  This work was supported in part by the National Science
Foundation and by the U.S. Department of Energy.

\eject\medskip\bigskip {\Large {\bf Figure captions}} \bigskip

Fig. 1. Detection of a free particle. The wave function of the detected
particle is non-negligible in the shaded cone.

Fig. 2. Detector placed under a potential barrier.

Fig. 3. Under-barrier detection. The wave function of the detected particle
is non-negligible in the shaded region, which is bounded by $R_{\max
}\left( x\right) $.

\end{document}